\documentclass{article}

\usepackage{arxiv}

\usepackage[utf8]{inputenc} 
\usepackage[T1]{fontenc} 
\usepackage{hyperref} 
\usepackage{url} 
\usepackage{booktabs} 
\usepackage{amsfonts} 
\usepackage{nicefrac} 
\usepackage{microtype} 
\usepackage{lipsum}
\usepackage{graphicx}
\usepackage{amsmath}
\usepackage{multirow}
\usepackage{xcolor} 
\usepackage{caption} 

\DeclareMathOperator{\logit}{logit}
\graphicspath{{./images/}{./figures/}}

\title{A Novel Exact Inference Approach for Log--Logistic Reliability Functions with Applications to Time-to-Event Data}

\author{ Bowen Liu \\
School of Science and Engineering\\
University of Missouri-Kansas City\\
Kansas City, MO 64110 \\
\texttt{bowen.liu@umkc.edu} \\
\And Malwane M.A. Ananda\\
Department of Mathematical Sciences\\
University of Nevada, Las Vegas\\
Las Vegas, NV 89154 \\
\texttt{malwane.ananda@unlv.edu} \\
\And Samaradasa Weerahandi\\
President of X-Techniques\\
Edison, NJ 08820\\
\texttt{weerahandi@aol.com} \\
}

\begin{document}
  \maketitle
  \begin{abstract}
    Log--logistic distribution is a flexible distribution that can model a wide range
    of failure patterns in the field of electrical, electronic and mechanical engineering and is
    often used in reliability inference. However, the inference of the parameters
    and reliability function of the log--logistic distribution can be challenging,
    especially in small sample scenarios. In this paper, we propose a new inference framework based on the least squares estimator–based generalized pivotal quantities (LSE-GPQ) for the parameters and reliability functions of the log--logistic distribution, which can provide better
    coverage in small sample scenarios. We will compare the performance of our
    proposed method with traditional methods such as the MLE and parametric
    bootstrapping through simulation studies and real data applications.
  \end{abstract}


  \section{Introduction}
  \label{sec:intro}

  In the field of electrical, electronic and mechanical engineering, reliability inference is
  of great importance for the design and maintenance of products \cite{meeker2021statistical,meeker1998accelerated,rekab2002sampling,rekab2013multistage,weerahandi1992testing,rausand2003system,lio2022bayesian,ananda1999estimation,liu2023new,liu2026exact}.
  The reliability of a product is defined as the probability that it will
  perform its intended function without failure for a specified period of time
  under stated conditions. Accurate reliability inference can help engineers identify
  potential failure modes, optimize maintenance schedules, and improve product design.
  In an era of rapid technological advancement, the demand for reliable products
  is increasing, making reliability inference an essential aspect of engineering.
  In the bio-medical field, such as in cancer research, the reliability inference is also known as the survival inference. In this context,
  the reliability of a treatment or intervention is often evaluated based on the
  time to an event of interest, such as the death or disease recurrence \cite{weerahandi2025approach,cooray2010analyzing,liu2023generalized,smith2017analysis,collett2023modelling,lio2022bayesian}.
  Accurate inference of survival data can help researchers identify effective
  treatments, optimize clinical trial design, and improve patient outcomes. In spite
  of its importance, the inference of reliability data or survival data can be
  challenging due to small sample sizes and censoring
  \cite{jia_exact_2018,jia_reliability_2015,weerahandi2025approach,balakrishnan_maximum_2008}.
  Therefore, developing robust and accurate methods for reliability inference is
  an active area of research in both engineering and medical fields.

  Parametric inference for reliability and survival problems is commonly used in
  engineering and medical research. These methods assume that the data follows a
  specific lifetime distribution, such as the exponential
  \cite{bain_statistical_2017,lawless_statistical_2011,cox_analysis_2018}, the Weibull
  \cite{rinne_weibull_2008,mudholkar_generalization_1996, mccool_using_2012} and the
  lognormal distribution
  \cite{chen_generalized_1995,kline_suitability_1984,panda_bootstrap-based_2025}.
  The log--logistic distribution is a flexible distribution that can model a wide
  range of failure rates and is often used in reliability inference \cite{gonzalez2026robust,tanis_record-based_2025,mohammed_computational_2025,gupta_study_1999}.
  Maximum likelihood estimation (MLE) is widely used for parameter estimation, with
  inference typically based on asymptotic properties, this approach may perform
  poorly in small sample scenarios. Similarly, methods such as parametric
  bootstrapping may not be reliable when the sample size is limited since they
  rely on the assumption that the estimated parameters are close to the true
  parameters. In such cases, the inference may be biased and have poor coverage
  probabilities.

  Therefore, developing robust methods for the statistical inference of the log--logistic distribution remains an important area of research in reliability inference. Inspired by the previous work on the exact inference for the quantities of Weibull distribution \cite{liu2026inference}, we propose a method based on generalized pivotal quantities (GPQs)
  of the parameters and reliability function of the log--logistic distribution,
  which can provide more accurate inference in small sample and censoring scenarios. We will compare the performance of our proposed method with traditional methods such as the MLE and the
  parametric bootstrapping through simulation studies and real data applications.

  \section{Preliminaries}
  \label{sec:prelim} In this section, we will introduce the concept of
  reliability inference and common techniques for reliability inference, focusing
  on parametric methods. We will also discuss the log--logistic distribution and
  its properties, which make it a useful distribution for modeling reliability data.
  Finally, we will review the challenges associated with inference for the log--logistic
  distribution, particularly in small sample and censoring scenarios.

  \subsection{Reliability Function}
  The reliability function, also known as the survival function, is a
  fundamental concept in reliability inference. It is defined as the probability
  that a product or system will survive beyond a certain time $t$.
  Mathematically, the reliability function can be expressed as:
  \begin{equation}
    R(t) = P(T > t) = 1 - F(t)
  \end{equation}
  where $T$ is the random variable representing the time to failure, and $F(t)$ is
  the cumulative distribution function (CDF) of $T$. The reliability function
  provides important information about the expected lifetime of a product or
  system and can be used to make informed decisions about maintenance and
  replacement schedules. Alternatively, in the context of survival analysis, the
  reliability function can be used to estimate the probability of survival for a
  patient or group of patients over time. (Usually denoted as $S(t)$ in survival
  analysis, but we will use $R(t)$ for consistency with reliability inference.)
  Assuming $T$ is a random variable associated with parameter vector $\theta$, the
  reliability function can be perceived as a function of $\theta$, i.e.,
  $R(t) = R(t; \theta)$. 

  \subsection{Asymptotic Inference}
  \label{sub:ai} Consider a random sample $T_{1}, T_{2}, ..., T_{n}$ of size $n$
  from a population with a continuous distribution function $F(t)$ and
  corresponding probability density function (PDF) $f(t)$. The likelihood function
  for the sample can be expressed as:
  \begin{equation}
    L(\theta) = \prod_{i=1}^{n}f(T_{i}; \theta)
  \end{equation}
  where $\theta$ is the parameter vector of the distribution. The maximum likelihood
  estimator (MLE) of $\theta$ is the value that maximizes the log-likelihood function,
  and can be found by solving the equation:
  \begin{equation}
    \frac{\partial l(\theta)}{\partial \theta}= 0
  \end{equation}
  Denote the MLE of $\theta$ as $\hat{\theta}$. Under appropriate regularity conditions, the
  MLE is asymptotically normal with mean $\theta$ and variance-covariance matrix
  given by the inverse of the Fisher information matrix, i.e.,
  \begin{equation}
    \sqrt{n}(\hat{\theta}- \theta) \xrightarrow{d}N(0, I^{-1}(\theta))
  \end{equation}
  where $I(\theta)$ is the Fisher information matrix and
  $I(\theta)= -E(\frac{\partial^{2}l}{\partial \theta^{2}})$. For a fixed value of
  $t$, the invariance property of the MLE gives the MLE for $R(t;\theta)$ as
  $R(t;\hat{\theta})$. The asymptotic variance can be estimated using the delta method
  as follows:
  \begin{equation}
    \sqrt{n}(\hat{R}(t; \theta)- R(t; \theta)) \xrightarrow{d}N(0, \nabla R(t; \theta
    )^{T}I^{-1}(\theta) \nabla R(t; \theta))
  \end{equation}
  where $\nabla R(t; \theta)$ is the gradient of $R(t; \theta)$ with respect to
  $\theta$. This result allows us to construct confidence intervals and perform hypothesis
  tests for $R(t)$ based on the asymptotic distribution of the MLE. However, in small
  sample scenarios, the asymptotic properties of the MLE may not hold, and the inference
  may be biased and have poor coverage probabilities. Therefore, alternative methods
  such as exact inference or resampling techniques may be necessary to obtain reliable
  results in such cases.

  \subsection{Parametric Bootstrapping}
  \label{sub:pb} Parametric bootstrapping also provides a way to construct confidence
  intervals for $R(t; \theta)$, which is based on the idea of resampling from
  the estimated distribution. The procedure involves the following steps:
  \begin{itemize}
    \item Fit the parametric model to the observed data and obtain the MLE $\hat{\theta}$.

    \item Generate a large number of bootstrap samples by simulating data from the
      fitted model using $\hat{\theta}$.

    \item For each bootstrap sample, calculate the MLE $\hat{\theta}^{*}$ and
      the corresponding estimate of $R(t; \hat{\theta}^{*})$.

    \item Use the distribution of the bootstrap estimates to construct confidence
      intervals for $R(t; \theta)$.
  \end{itemize}
  While parametric bootstrapping can be a useful tool for inference, it relies on
  the assumption that the estimated parameters are close to the true parameters.
  In small sample scenarios, this assumption may not hold, and the inference may
  be biased and have poor coverage probabilities. Therefore, it is essential to
  consider alternative methods, such as exact inference or non-parametric bootstrapping,
  when dealing with small sample sizes.

  \subsection{log--logistic distribution}
  The log--logistic distribution is a continuous probability distribution that is
  often used to model reliability data and survival data. Suppose $T$ is a random
  variable that follows a log--logistic distribution with shape parameter $\alpha>
  0$ and scale parameter $\beta>0$, denoted as
  $T \sim \mathrm{log--logistic}(\alpha,\beta)$. The probability density function (PDF)
  of $T$ is given by
  \begin{equation}
    \label{eqn:llpdf}f(t;\alpha,\beta) =\frac{\alpha}{\beta}\left(\frac{t}{\beta}
    \right)^{\alpha-1}\left[1+\left(\frac{t}{\beta}\right)^{\alpha}\right]^{-2},\quad
    t>0.
  \end{equation}
  The cumulative distribution function (CDF) of $T$ is
  \begin{equation}
    \label{eqn:llcdf}F(t;\alpha,\beta) =\frac{1}{1+\left(\frac{t}{\beta}\right)^{-\alpha}}
    =\frac{\left(\frac{t}{\beta}\right)^{\alpha}}{1+\left(\frac{t}{\beta}\right)^{\alpha}}
    ,\quad t>0.
  \end{equation}
  The reliability function of $T$ can be expressed as
  \begin{equation}
    \label{eqn:llrel}R(t;\alpha,\beta) =1-F(t;\alpha,\beta) =\frac{1}{1+\left(\frac{t}{\beta}\right)^{\alpha}}
    ,\quad t>0.
  \end{equation}

\begin{figure}[!htbp]
    \centering
    \includegraphics[width=0.8\textwidth]{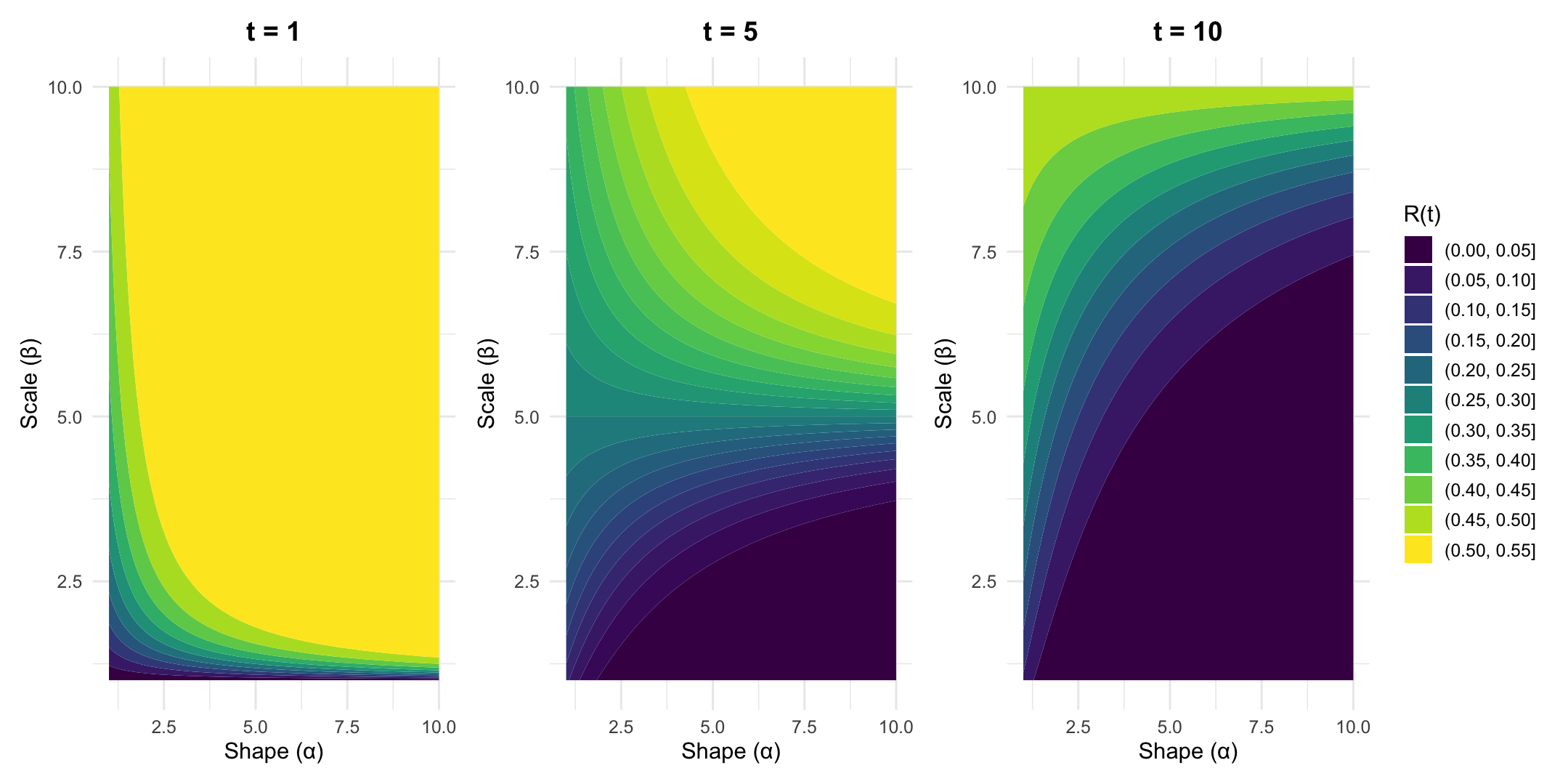}
    \caption{Contour Plot of Reliability Function of log--logistic Distribution
    with Different Parameters at $t=1, 5, \text{ and }10$.}
    \label{fig:figure1}
  \end{figure}

  Thus, when $t$ is fixed, $R(t; \alpha, \beta)$ can be perceived as a function
  of $\alpha$ and $\beta$. Figure \ref{fig:figure1} shows the contour plot of the
  reliability function of the log--logistic distribution with different choices of
  parameters at $t=1, 5, \text{ and }10$. Essentially, with the methods in \ref{sub:ai}
  and \ref{sub:pb}, we can construct confidence intervals for
  $R(t; \alpha, \beta)$ based on the asymptotic distribution of the MLE or the
  distribution of the bootstrap estimates. Specifically, the MLE of $\alpha$ and
  $\beta$ can be obtained by maximizing the loglikelihood function based on the
  PDF in \ref{eqn:llpdf}. The loglikelihood function can be expressed as:
  \begin{equation}
    \ell(\alpha,\beta) = \sum_{i=1}^{n}\log f(T_{i};\alpha,\beta) = \sum_{i=1}^{n}
    \left[ \log\alpha -\log\beta +(\alpha-1)\log\!\left(\frac{T_{i}}{\beta}\right
    ) -2\log\!\left(1+\left(\frac{T_{i}}{\beta}\right)^{\alpha}\right) \right].
  \end{equation}

  Denote the MLE of $\alpha$ and $\beta$ as $\hat{\alpha}$ and $\hat{\beta}$,
  respectively. For fixed $t$, the MLE of $R(t; \alpha, \beta)$ can be obtained
  by plugging in the MLEs of $\alpha$ and $\beta$, i.e., $\hat{R}(t; \alpha, \beta
  ) = R(t; \hat{\alpha}, \hat{\beta})$. The confidence intervals for
  $R(t; \alpha, \beta)$ can be constructed based on the asymptotic distribution
  of the MLE or the distribution of the bootstrap estimates.

  However, as mentioned earlier, these methods may not perform well in small
  sample scenarios, and alternative inference methods such based on generalized pivotal
  quantities (GPQ) may be necessary to obtain reliable results with better coverage
  probabilities.

  \section{Methodology}
  \label{sec:method}
  \subsection{Least Square Estimator (LSE) for log--logistic Distribution}
  Though parameter estimation based on likelihood function with PDF in
  \ref{eqn:llpdf} is widely used, it may not perform well in small sample scenarios with Type-I censored data.
  Therefore, we propose a LSE (Least Square Estimator ) for the parameters of
  the log--logistic distribution. To illustrate it, we first introduce the
  concept of LSE method for parameter estimation of log--logistic distribution. This
  is motivated by Zhang's work on the LSE for the Weibull distribution \cite{zhang_study_2007}.
  Previously, researchers have conducted studies on the GPQs for the parameters of
  Weibull distribution based on LSE method \cite{jia_exact_2018,jia_reliability_2015,liu2026inference}
  and have shown that the GPQs based on LSE method can provide more accurate inference
  in small sample scenarios. In this paper, we extend the LSE method to the parameter
  estimation of log--logistic distribution.

  Suppose $T_{1}, ..., T_{n}$ are i.i.d random variables from a log--logistic distribution
  with shape parameter $\alpha$ and scale parameter $\beta$. Then, for each $i$,
  we have
  \begin{equation}
    F(T_{i}; \alpha, \beta) = \frac{1}{1+(T_{i}/\beta)^{-\alpha}}
  \end{equation}

  Applying the logarithmic transformation $Y_{i}=\log T_{i}$ converts the log--logistic
  random variable to a logistic location--scale random variable. Specifically,
  we can represent $Y_{i}$ as
  \[
    Y_{i}=\mu+sZ_{i},
  \]
  where $Z_{i}\sim\mathrm{Logistic}(0,1)$, $\mu=\log\beta$, and $s=1/\alpha$.

  Therefore, the CDF of $Z_{i}$ can be expressed as
  \begin{equation}
    F_{Z_i}(z_{i}) = \frac{1}{1+e^{-z_i}}.
  \end{equation}

  And, we can obtain the inverse CDF of $Z_{i}$ as
  \begin{equation}
    F_{Z_i}^{-1}(p_{i}) = \logit(p_{i}) = \log\left(\frac{p_{i}}{1-p_{i}}\right)
  \end{equation}

  Let $p_{i}$ denote plotting positions (for complete data, Benard's
  approximation $p_{i}=(i-0.3)/(n+0.4)$ is commonly used), and define the
  logistic probability--plot regressors $x_{i}=\mathrm{logit}(p_{i})=\log\{p_{i}/
  (1-p_{i})\}$.

  The LSEs $\hat{s}$ and $\hat{\mu}$ are obtained by fitting the linear regression
  $y_{(i)}= \mu+s x_{i}$, where $\hat{s}$ is the slope, $\hat{\mu}$ the intercept,
  and $y_{(i)}$ represents the $i$-th ordered sample point in . Specifically, the
  LSEs can be calculated as
  \begin{equation}
    \begin{aligned}
      \hat{s}   & = \frac{\sum_{i=1}^{n}(x_{i}-\bar{x})(y_{(i)}-\bar{y})}{\sum_{i=1}^{n}(x_{i}-\bar{x})^{2}}, \\
      \hat{\mu} & = \bar{y} - \hat{s}\,\bar{x}.
    \end{aligned}
  \end{equation}
  where $\bar{x}$ and $\bar{y}$ are the sample means of $x_{i}$ and $y_{(i)}$,
  respectively. The LSEs $\hat{\alpha}$ and $\hat{\beta}$ can be obtained by mapping
  back to the original parameters, i.e., $\hat{\alpha}=1/\hat{s}$ and
  $\hat{\beta}=\exp(\hat{\mu})$.

    \subsection{Censored Data Least Squares Estimation}

When the data are subject to right censoring, the least-squares estimation
procedure must be modified to account for the incomplete observations. As in
the complete-data case, we work with the transformed variables
$Y_i = \log T_i$, which follow a logistic location--scale model
\[
Y_{(i)} = \mu + s x_i, \qquad \mu = \log \beta, \qquad s = \frac{1}{\alpha}.
\]

However, under censoring, the plotting positions $p_i$ are no longer obtained
from fixed formulas such as Benard's approximation. Instead, they are estimated
from the data using a nonparametric estimator of the distribution function,
such as the Kaplan--Meier (KM) estimator.

Let
\[
\hat{p}_i = \hat{F}_Y\bigl(t_{(i)}\bigr)
\]
denote the estimated distribution function evaluated at the ordered
observations $t_{(i)}$. For the uncensored (failure) observations, we define
the regressors
\[
x_i = \logit(\hat{p}_i) = \log\left(\frac{\hat{p}_i}{1-\hat{p}_i}\right).
\]

Let $\mathcal{F}$ denote the set of indices corresponding to failure
observations, and define $y_i = \log t_{(i)}$. The least-squares estimators
based on censored data are then obtained by fitting the regression model
over the failure set $\mathcal{F}$:
\[
\hat{s}
=
\frac{\sum\limits_{i \in \mathcal{F}} (x_i - \bar{x})(y_i - \bar{y})}
{\sum\limits_{i \in \mathcal{F}} (x_i - \bar{x})^2},
\qquad
\hat{\mu}
=
\bar{y} - \hat{s}\,\bar{x},
\]
where $\bar{x}$ and $\bar{y}$ are the sample means computed over the failure
indices $\mathcal{F}$.

The estimators of the original parameters are then given by
\[
\hat{\alpha} = \frac{1}{\hat{s}}, \qquad \hat{\beta} = \exp(\hat{\mu}).
\]
  \subsection{Generalized Pivotal Quantities (GPQ) of Parameters}

  To construct generalized pivotal quantities for the parameters of the log--logistic
  distribution, we exploit the location--scale representation $Y=\mu+sZ$, where
  $Z\sim\mathrm{Logistic}(0,1)$ and the plotting-position regressors
  $x_{i}=\operatorname{logit}(p_{i})$ are fixed and free of unknown parameters. Let
  $\hat{s}$ denote the least squares estimator obtained from the observed data by
  regressing the ordered responses $y_{(i)}$ on $x_{i}$. Under the model
  $Y_{(i)}=\mu+sZ_{(i)}$, the LSE slope has the representation as follows:
  \[
    \hat{s}= \frac{\sum_{i=1}^{n}(x_{i}-\bar{x})(y_{(i)}-\bar{y})}{\sum_{i=1}^{n}(x_{i}-\bar{x})^{2}}
    = s\,\frac{\sum_{i=1}^{n}(x_{i}-\bar{x})(Z_{(i)}-\bar{Z})}{\sum_{i=1}^{n}(x_{i}-\bar{x})^{2}}
    = s\,\hat{s}(Z),
  \]
  where $\hat{s}(Z)$ is the random slope obtained by regressing the ordered
  standard logistic sample $Z_{(i)}$ on the same fixed regressors $x_{i}$. Let $\tilde
  {s}$ denote the random variable representing $\hat{s}$ in repeated sampling.
  The above relation implies $\tilde{s}=s\,\hat{s}(Z)$, since the location parameter
  $\mu$ cancels out in the slope.

  Consequently, the ratio
  \[
    G_{s}=\frac{\hat{s}}{\hat{s}(Z)}
  \]
  is a generalized pivotal quantity for $s$, because it satisfies the defining properties
  of a GPQ: (i) at the observed sample points, $y_{(i)}=\mu+sZ_{(i)}$ implies $\hat
  {s}=s\,\hat{s}(Z)$ and hence $G_{s}=s$; and (ii) the distribution of
  $\hat{s}(Z)$, and therefore of $G_{s}$, depends only on the standard logistic
  distribution and the fixed design points $x_{i}$, and is free of unknown parameters.

  A GPQ for the location parameter $\mu$ is derived from the regression identity
  $\hat{\mu}=\bar{y}-\hat{s}\,\bar{x}$. Writing this identity in terms of the
  standardized representation yields
  \[
    \hat{\mu}= \mu+s\bar{Z}-s\,\bar{x}\,\hat{s}(Z),
  \]
  and hence the random version can be rearranged as
  \[
    \mu=\hat{\mu}-s\{\bar{Z}-\bar{x}\,\hat{s}(Z)\}.
  \]
  Replacing $s$ by its GPQ $G_{s}$ leads to the generalized pivotal quantity
  \[
    G_{\mu}=\hat{\mu}-G_{s}\{\bar{Z}-\bar{x}\,\hat{s}(Z)\},
  \]
  which reduces to $\mu$ at the observed data and whose distribution is free of
  unknown parameters.

  Finally, mapping back to the original log--logistic parameters yields
  \[
    G_{\alpha}=\frac{1}{G_{s}}, \qquad G_{\beta}=\exp(G_{\mu}).
  \]
  Using the substitution principle for generalized inference, a GPQ for the
  reliability function $R(t;\alpha,\beta)=\{1+(t/\beta)^{\alpha}\}^{-1}$ is
  obtained as
  \[
    G_{R}(t)=\left\{1+\left(\frac{t}{G_{\beta}}\right)^{G_\alpha}\right\}^{-1},
  \]
  and generalized confidence limits for $R(t;\alpha,\beta)$ are constructed from
  the empirical quantiles of $G_{R}(t)$.

\begin{figure}[!htbp]
    \centering
    \includegraphics[width=0.8\textwidth]{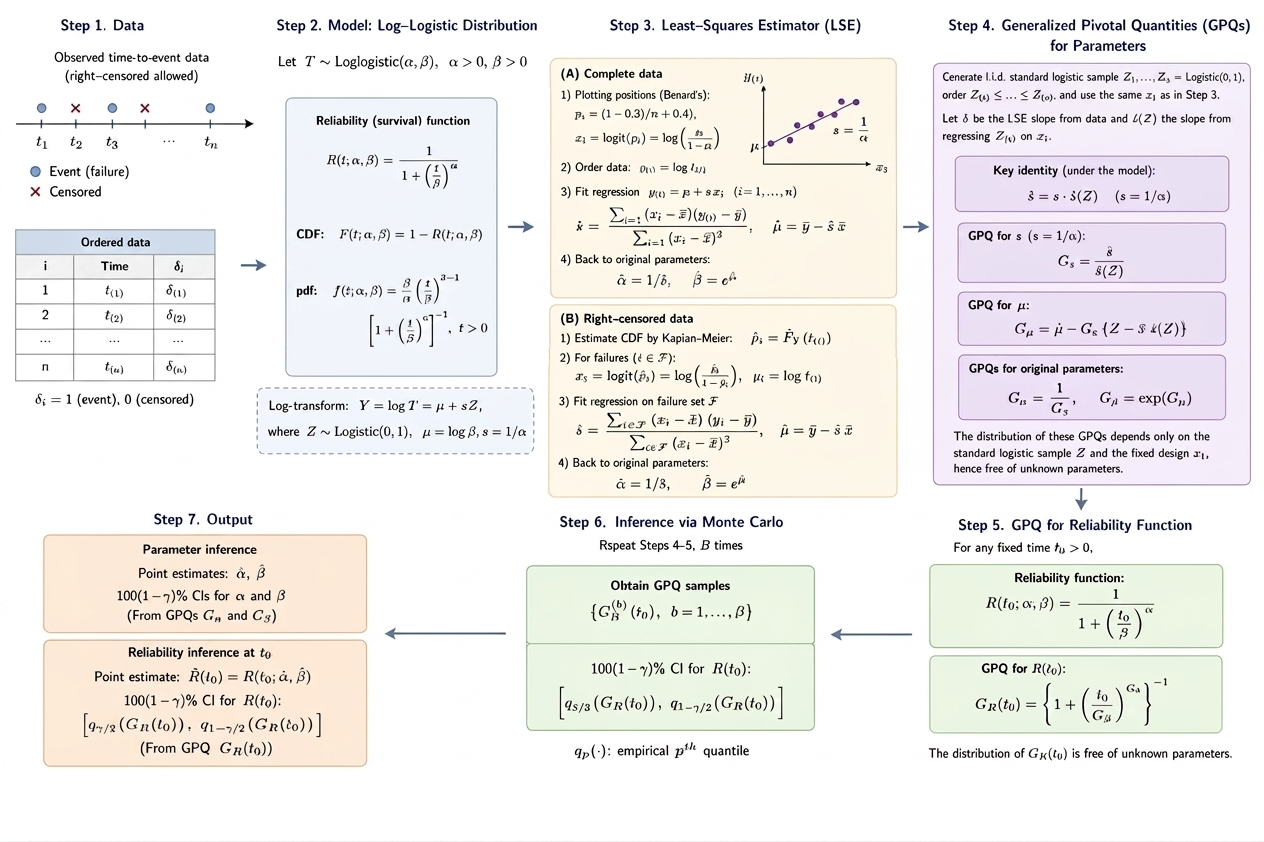}
    \caption{Framework of LSE-GPQ Method for Inference of Quantities of log--logistic Distribution.}
    \label{fig:figure2}
    \end{figure}

  The framework of LSE-GPQ method is shown in Figure \ref{fig:figure2}, where the LSEs are obtained by fitting the regression model to the observed data, and the GPQs are constructed based on the LSEs obtained from the observed data and the standard logistic random variables.  

  \section{Simulations}

  \subsection{Complete Data}
  To illustrate the performance of the proposed LSE-GPQ method, we conduct limited
  simulation studies to compare it with traditional methods such as MLE and parametric
  bootstrapping. We consider different sample sizes ($n=10$ and $n=20$), different
  values of $t$ (1 and 2), and different combinations of shape ($\alpha = 2 \text{
  and }5$) and scale parameters ($\beta = 1 \text{ and }2$) for the log--logistic
  distribution.

  For each simulation scenario, we generated $r = 1,000$ from the log--logistic distribution
  with specified parameters and calculate the confidence intervals for $R(t; \alpha
  , \beta)$ using the LSE-GPQ method, MLE-based asymptotic inference, and parametric
  bootstrapping. The coverage probabilities will be estimated as the proportion
  of confidence intervals that contain the true value of $R(t; \alpha, \beta)$. For
  number of bootstrap samples, we set it to be 2,000 for each simulation
  scenario. For LSE-GPQ method, we generate 2,000 samples of $G_{R}(t)$ for each
  replicate under each scenario to construct the confidence intervals.

  The results of the simulation studies are presented in table
  \ref{tab:coverage_090} and \ref{tab:coverage_095}, which shows the coverage probabilities
  at the nominal level of 0.90 and 0.95 for different methods under various scenarios.
  The LSE-GPQ method consistently provides better coverage probabilities
  compared to MLE and parametric bootstrapping when sample size is small. This demonstrates
  the advantage of the proposed method in providing more accurate inference for the
  reliability function of the log--logistic distribution.

  \begin{center}
    \captionof{table}{Coverage probabilities at nominal level 0.90} \label{tab:coverage_090}

    \begin{tabular}{ccccccc}
      \toprule \multicolumn{7}{c}{Nominal Level: 0.90} \\
      \midrule $n$                                    & $t$                & Shape & Scale & \multicolumn{3}{c}{Coverage Probability} \\
      \cmidrule(lr){5-7}                              &                    &       &       & LSE--GPQ                                & PB    & AI    \\
      \midrule \multirow{8}{*}{10}                    & \multirow{4}{*}{1} & 2     & 1     & 0.901                                   & 0.857 & 0.841 \\
                                                      &                    & 2     & 2     & 0.907                                   & 0.846 & 0.815 \\
                                                      &                    & 5     & 1     & 0.901                                   & 0.857 & 0.841 \\
                                                      &                    & 5     & 2     & 0.904                                   & 0.838 & 0.757 \\
      \cmidrule(lr){2-7}                              & \multirow{4}{*}{2} & 2     & 1     & 0.914                                   & 0.834 & 0.802 \\
                                                      &                    & 2     & 2     & 0.901                                   & 0.857 & 0.841 \\
                                                      &                    & 5     & 1     & 0.911                                   & 0.818 & 0.735 \\
                                                      &                    & 5     & 2     & 0.901                                   & 0.857 & 0.841 \\
      \midrule \multirow{8}{*}{20}                    & \multirow{4}{*}{1} & 2     & 1     & 0.903                                   & 0.873 & 0.867 \\
                                                      &                    & 2     & 2     & 0.902                                   & 0.867 & 0.853 \\
                                                      &                    & 5     & 1     & 0.889                                   & 0.845 & 0.722 \\
                                                      &                    & 5     & 2     & 0.886                                   & 0.844 & 0.809 \\
      \cmidrule(lr){2-7}                              & \multirow{4}{*}{2} & 2     & 1     & 0.901                                   & 0.873 & 0.862 \\
                                                      &                    & 2     & 2     & 0.903                                   & 0.873 & 0.867 \\
                                                      &                    & 5     & 1     & 0.894                                   & 0.853 & 0.809 \\
                                                      &                    & 5     & 2     & 0.903                                   & 0.873 & 0.867 \\
      \bottomrule
    \end{tabular}
  \end{center}

  \begin{center}
    \captionof{table}{Coverage probabilities at nominal level 0.95} \label{tab:coverage_095}

    \begin{tabular}{ccccccc}
      \toprule \multicolumn{7}{c}{Nominal Level: 0.95} \\
      \midrule $n$                                    & $t$                & Shape & Scale & \multicolumn{3}{c}{Coverage Probability} \\
      \cmidrule(lr){5-7}                              &                    &       &       & LSE--GPQ                                & PB    & AI    \\
      \midrule \multirow{8}{*}{10}                    & \multirow{4}{*}{1} & 2     & 1     & 0.943                                   & 0.912 & 0.893 \\
                                                      &                    & 2     & 2     & 0.953                                   & 0.904 & 0.862 \\
                                                      &                    & 5     & 1     & 0.943                                   & 0.912 & 0.893 \\
                                                      &                    & 5     & 2     & 0.960                                   & 0.881 & 0.660 \\
      \cmidrule(lr){2-7}                              & \multirow{4}{*}{2} & 2     & 1     & 0.952                                   & 0.897 & 0.846 \\
                                                      &                    & 2     & 2     & 0.943                                   & 0.912 & 0.893 \\
                                                      &                    & 5     & 1     & 0.957                                   & 0.878 & 0.766 \\
                                                      &                    & 5     & 2     & 0.943                                   & 0.912 & 0.893 \\
      \midrule \multirow{8}{*}{20}                    & \multirow{4}{*}{1} & 2     & 1     & 0.947                                   & 0.930 & 0.922 \\
                                                      &                    & 2     & 2     & 0.955                                   & 0.923 & 0.907 \\
                                                      &                    & 5     & 1     & 0.947                                   & 0.930 & 0.922 \\
                                                      &                    & 5     & 2     & 0.944                                   & 0.907 & 0.836 \\
      \cmidrule(lr){2-7}                              & \multirow{4}{*}{2} & 2     & 1     & 0.950                                   & 0.919 & 0.897 \\
                                                      &                    & 2     & 2     & 0.947                                   & 0.930 & 0.922 \\
                                                      &                    & 5     & 1     & 0.943                                   & 0.907 & 0.844 \\
                                                      &                    & 5     & 2     & 0.947                                   & 0.930 & 0.922 \\
      \bottomrule
    \end{tabular}
  \end{center}

\subsection{Censoring Data}

To evaluate the performance of the proposed method in the presence of censoring, we conduct additional simulation studies under various type-1 censoring scenarios. We consider sample size ($n$) of $10$, different combinations of shape ($\alpha = 2 \text{ and }5$) and scale parameters ($\beta = 1 \text{ and }2$) for the log--logistic distribution, and different censoring proportions (20\% and 50\%).
The results are shown in table \ref{tab:coverage_censoring}. The proposed LSE-GPQ method still provides better coverage probabilities compared to MLE and parametric bootstrapping in the presence of censoring, demonstrating its robustness and effectiveness across all type-1 censored data scenarios.

\begin{center}
\captionof{table}{Coverage probabilities at nominal level 0.95}
\label{tab:coverage_censoring}

\begin{tabular}{ccccccc}
\toprule
\multicolumn{7}{c}{Nominal Level: 0.95} \\
\midrule
$t$ & Shape & Scale & Censoring proportion & \multicolumn{3}{c}{Coverage Probability} \\
\cmidrule(lr){5-7}
    &       &       &                      & LSE--GPQ & PB & AI \\
\midrule

\multirow{8}{*}{1} 
& 2 & 1 & 20\% & 0.941 & 0.911 & 0.916 \\
& 2 & 1 & 50\% & 0.948 & 0.927 & 0.907 \\
& 2 & 2 & 20\% & 0.952 & 0.939 & 0.909 \\
& 2 & 2 & 50\% & 0.937 & 0.918 & 0.904 \\
& 5 & 1 & 20\% & 0.944 & 0.917 & 0.912 \\
& 5 & 1 & 50\% & 0.945 & 0.915 & 0.893 \\
& 5 & 2 & 20\% & 0.947 & 0.931 & 0.842 \\
& 5 & 2 & 50\% & 0.953 & 0.894 & 0.769 \\
\cmidrule(lr){1-7}
\multirow{8}{*}{2} 
& 2 & 1 & 20\% & 0.941 & 0.915 & 0.896 \\
& 2 & 1 & 50\% & 0.939 & 0.935 & 0.869 \\
& 2 & 2 & 20\% & 0.949 & 0.930 & 0.927 \\
& 2 & 2 & 50\% & 0.937 & 0.918 & 0.904 \\
& 5 & 1 & 20\% & 0.946 & 0.913 & 0.829 \\
& 5 & 1 & 50\% & 0.940 & 0.945 & 0.825 \\
& 5 & 2 & 20\% & 0.950 & 0.935 & 0.935 \\
& 5 & 2 & 50\% & 0.941 & 0.901 & 0.878 \\
\bottomrule
\end{tabular}
\end{center}

We envision the similar performance of the proposed method in the presence multiply type 1 censoring or type 2 censoring, which is common in reliability
  inference and survival analysis. Also, based on the invariance property
  of GPQs, the GPQs for the parameters of the log--logistic distribution can be
  easily mapped to the GPQ for any functions of the log--logistic parameters
  $\alpha$ and $\beta$, which allows us to construct confidence intervals for other
  quantities of interest, such as the mean time between failures or the median
  time between failures, in addition to the reliability function.

  \section{Real Data Applications}

  In this section, we apply the proposed LSE-GPQ method to real data examples to
  demonstrate its practical utility. We will apply the method to three different
  data sets: (1) failure time of grinders, (2) secondary reactor pump data, and (3)
  electrical breakdown of an insulating fluid data. For each data set, we will first
  provide a histogram of the observed data along with the fitted density curves based
  on MLE and LSE. We will also present a table summarizing the data set and
  conduct goodness-of-fit tests to justify the log--logistic fitting. Specifically,
  we will use the Kolmogorov-Smirnov (KS) test to assess the goodness-of-fit of
  the log--logistic distribution to the observed data. The KS test statistic is defined
  as follows:
  \begin{equation}
    D = \sup_{t}|F_{n}(t) - F(t; \hat{\alpha}, \hat{\beta})|
  \end{equation}
  where $F_{n}(t)$ is the empirical distribution function of the observed data, and
  $F(t; \hat{\alpha}, \hat{\beta})$ is the CDF of the log--logistic distribution with
  parameters estimated by MLE or LSE. P-values for KS test are calculated based
  on the asymptotic distribution of the test statistics under the null
  hypothesis.

  The null hypothesis for KS test is that the data follows a log--logistic
  distribution, and the alternative hypothesis is that the data does not follow a
  log--logistic distribution.

  Finally, we will compare the average length of confidence intervals for
  $R(t; \alpha, \beta)$ at different values of $t$ using the proposed LSE-GPQ
  method and traditional methods.

  \subsection{Failure Time of Grinders}
\begin{figure}[!htbp]
    \centering
    \includegraphics[width=0.6\textwidth]{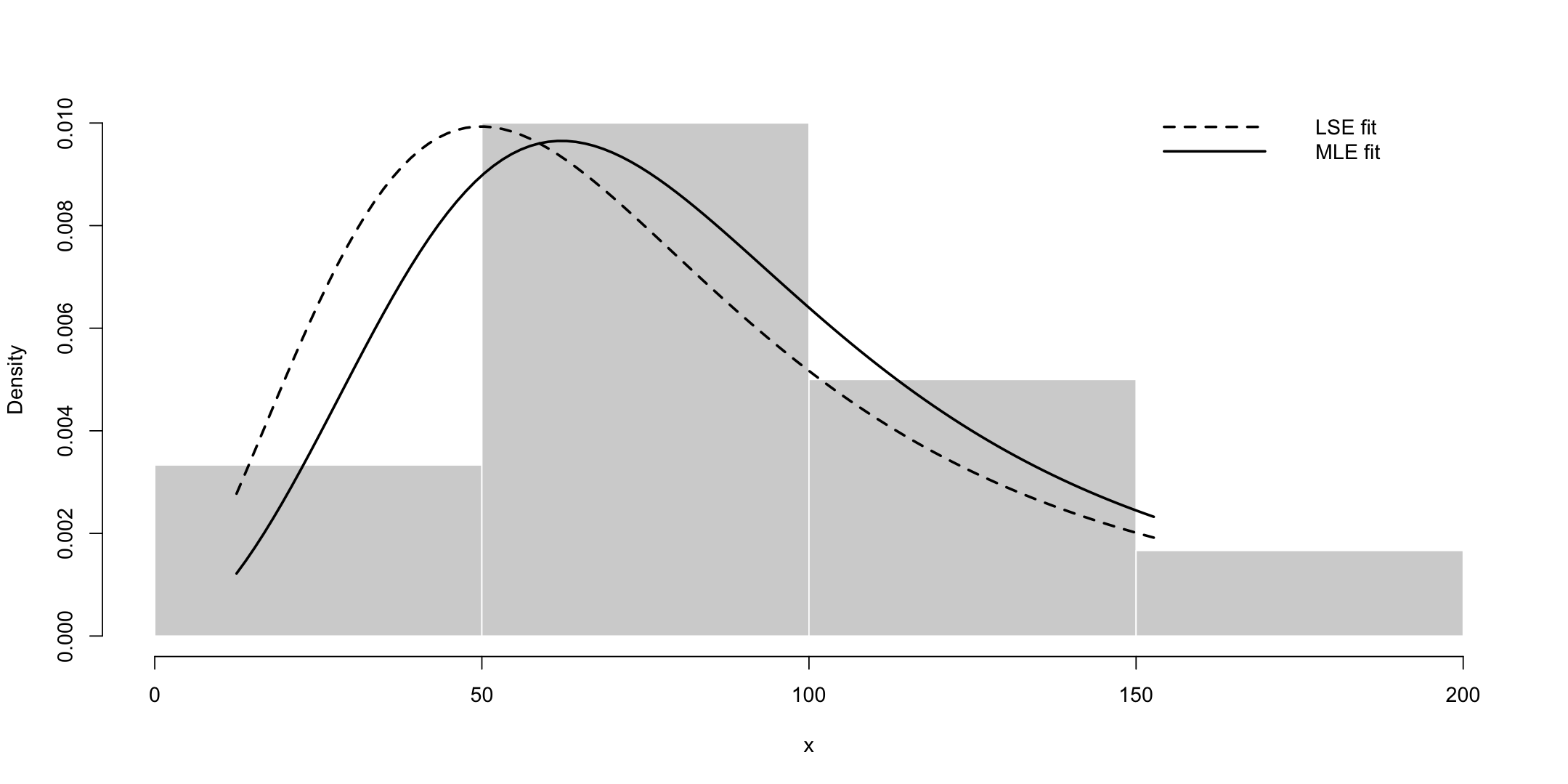}
    \caption{Histogram of the Grinder Data with LSE fit and MLE fit.}
    \label{fig:grinder_hist}
  \end{figure}

  The data set contains the observed failure times of 12 grinders and was
  previously used by researchers for reliability inference with Weibull
  distribution \cite{jia_reliability_2015, balakrishnan_maximum_2008}. The summary
  statistics of the data set are presented in Table \ref{tab:summary_grinders}.
  The histogram of the data set is shown in Figure \ref{fig:grinder_hist}, along
  with the fitted density curves based on MLE and LSE.

  We have performed the KS test and AD test with both MLE and LSE to assess the goodness-of-fit
  of the log--logistic distribution to the observed data (LSE: $D = 0.245, \text{p-value}
  = 0.403$; MLE: $D = 0.189, \text{p-value}= 0.721$). The p-values for both tests
  are greater than 0.05, indicating that we fail to reject the null hypothesis
  and suggesting that the log--logistic distribution is a reasonable fit for the
  data.

  \begin{center}
    \captionof{table}{Summary Statistics of the Failure Time of the Grinders Data}
    \label{tab:summary_grinders}

    \begin{tabular}{lcccccc}
      \toprule Statistic & Min   & 1st Quartile & Median & Mean  & 3rd Quartile & Max    \\
      \midrule Value     & 12.50 & 65.55        & 96.05  & 86.42 & 116.45       & 152.70 \\
      \bottomrule
    \end{tabular}
  \end{center}

  To evaluate the performance of the proposed LSE-GPQ method against traditional
  methods, we have calculated the average length of confidence intervals for $R(t
  ; \alpha, \beta)$ at different values of $t$ ($t = \text{1st quartile, median,
  3rd quartile}$) using the LSE-GPQ method, MLE-based asymptotic inference, and parametric
  bootstrapping. The results are presented in Table \ref{tab:ci_length_grinders}.
  The LSE-GPQ method provides shorter confidence intervals compared to MLE and parametric
  bootstrapping, demonstrating its advantage in providing more accurate inference for the reliability function of the log--logistic distribution.

  \begin{center}
    \captionof{table}{Empirical reliability and 95\% confidence intervals at selected time points for the grinder Data}
    \label{tab:ci_length_grinders}

    \begin{tabular}{lccccc}
      \toprule $t$                                    & Empirical $R(t)$       & Method   & 95\% CI        & Interval Length \\
      \midrule \multirow{3}{*}{1st Quartile = 65.55}  & \multirow{3}{*}{0.750} & LSE--GPQ & (0.315, 0.789) & 0.474           \\
                                                      &                        & PB       & (0.395, 0.872) & 0.477           \\
                                                      &                        & AI       & (0.409, 0.888) & 0.478           \\
      \midrule \multirow{3}{*}{Median = 96.05}        & \multirow{3}{*}{0.500} & LSE--GPQ & (0.136, 0.579) & 0.443           \\
                                                      &                        & PB       & (0.144, 0.648) & 0.503           \\
                                                      &                        & AI       & (0.164, 0.618) & 0.455           \\
      \midrule \multirow{3}{*}{3rd Quartile = 116.45} & \multirow{3}{*}{0.250} & LSE--GPQ & (0.079, 0.472) & 0.393           \\
                                                      &                        & PB       & (0.071, 0.502) & 0.434           \\
                                                      &                        & AI       & (0.073, 0.475) & 0.402           \\
      \bottomrule
    \end{tabular}
  \end{center}

  \subsection{Secondary Reactor Pump Data}

\begin{figure}[!htbp]
    \centering
    \includegraphics[width=0.6\textwidth]{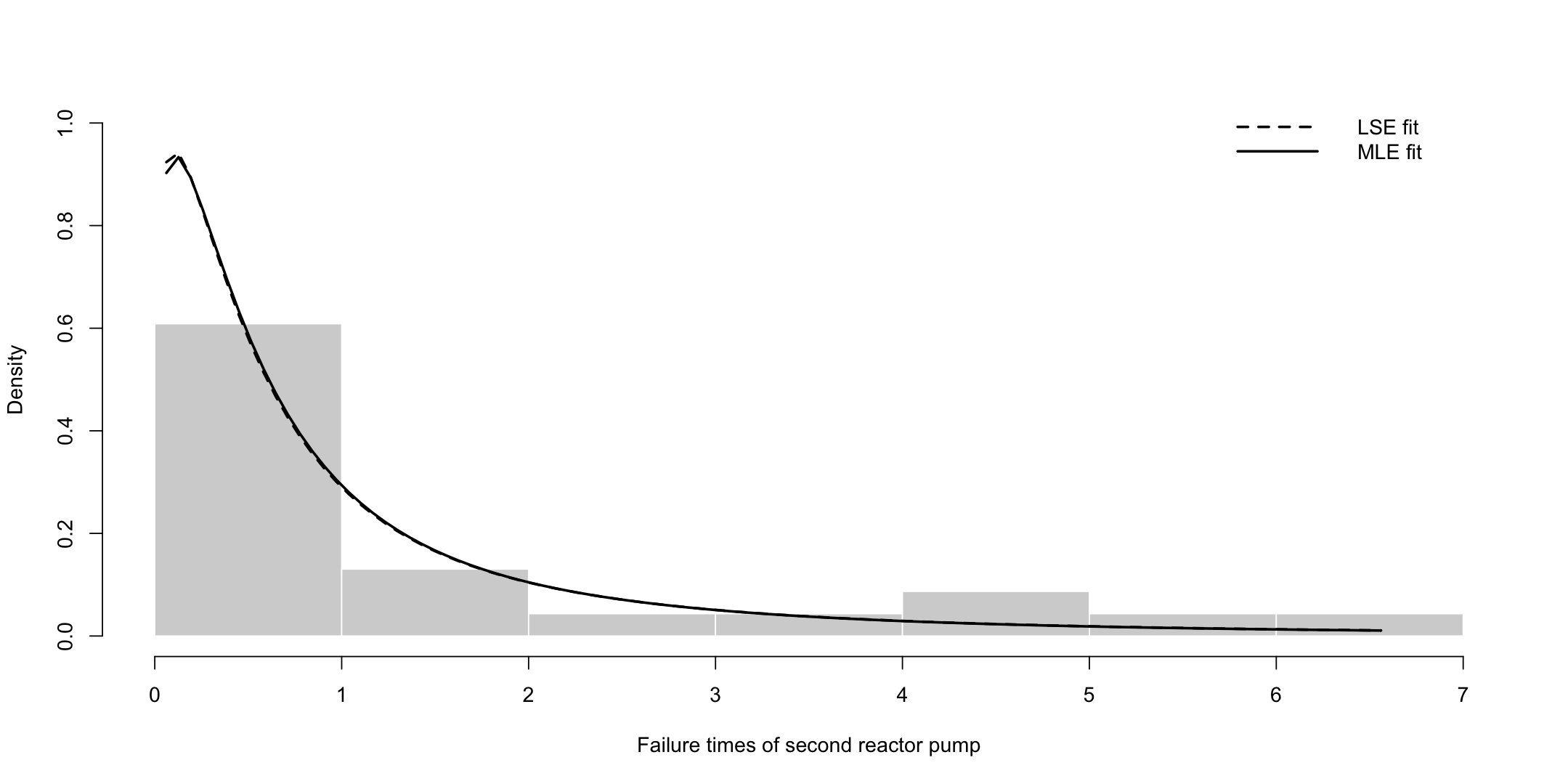}
    \caption{Histogram of the secondary reactor pumps data with LSE fit and MLE
    fit.}
    \label{fig:ifluid_hist}
  \end{figure}

  We also analyzed a data set of times between failures for secondary reactor
  pumps \cite{suprawhardana1999total}. Summary statistics of the data is presented
  in Table~\ref{tab:summary_reactor2}. This data set has been utilized in
  previous research on reliability modeling with heavy-tailed distributions
  \cite{liu2023new, sharma2014new}. The histogram of the data set is shown in Figure
  \ref{fig:ifluid_hist}, along with the fitted density curves based on MLE and
  LSE. We have performed the KS test and AD test with both MLE and LSE to assess
  the GoF of the log--logistic distribution to the observed data (LSE: $D = 0.093,
  \text{p-value}= 0.0.979$; MLE: $D = 0.090, \text{p-value}= 0.984$). The p-values
  for both tests are greater than 0.05, indicating that we fail to reject the
  null hypothesis and suggesting that the log--logistic distribution is a
  reasonable fit for the data.

  \begin{center}
    \captionof{table}{Summary statistics of the second reactor pump failure time data}
    \label{tab:summary_reactor2}

    \begin{tabular}{lcccccc}
      \toprule Statistic & Min   & 1st Quartile & Median & Mean  & 3rd Quartile & Max   \\
      \midrule Value     & 0.062 & 0.310        & 0.614  & 1.578 & 2.041        & 6.560 \\
      \bottomrule
    \end{tabular}
  \end{center}

  Similar to the previous data set, we have calculated the average length of confidence
  intervals for $R(t; \alpha, \beta)$ at different values of $t$ ($t = \text{1st
  quartile, median, 3rd quartile}$) using the LSE-GPQ method, MLE-based
  asymptotic inference, and bootstrapping based on MLE. The results are presented
  in Table \ref{tab:empirical_reliability_reactor2}. The LSE-GPQ method provides
  shorter confidence intervals compared to MLE and parametric bootstrapping,
  demonstrating its advantage in providing more accurate inference for the
  reliability function of the log--logistic distribution.

  \begin{center}
    \captionof{table}{Empirical reliability and 95\% confidence intervals at selected time points for the second reactor pump}
    \label{tab:empirical_reliability_reactor2}

    \begin{tabular}{lccccc}
      \toprule $t$                                   & Empirical $R(t)$       & Method   & 95\% CI        & Interval Length \\
      \midrule \multirow{3}{*}{1st Quartile = 0.310} & \multirow{3}{*}{0.739} & LSE--GPQ & (0.566, 0.868) & 0.302           \\
                                                     &                        & PB       & (0.569, 0.885) & 0.316           \\
                                                     &                        & AI       & (0.578, 0.889) & 0.311           \\
      \midrule \multirow{3}{*}{Median = 0.614}       & \multirow{3}{*}{0.478} & LSE--GPQ & (0.369, 0.709) & 0.340           \\
                                                     &                        & PB       & (0.352, 0.725) & 0.373           \\
                                                     &                        & AI       & (0.363, 0.723) & 0.360           \\
      \midrule \multirow{3}{*}{3rd Quartile = 2.041} & \multirow{3}{*}{0.261} & LSE--GPQ & (0.096, 0.374) & 0.278           \\
                                                     &                        & PB       & (0.083, 0.363) & 0.280           \\
                                                     &                        & AI       & (0.071, 0.356) & 0.285           \\
      \bottomrule
    \end{tabular}
  \end{center}

  \section{Concluding Remarks}

  In this paper, we have proposed a novel method for constructing confidence intervals
  for the reliability function of the log--logistic distribution based on generalized
  pivotal quantities (GPQ) derived from least square estimators (LSE). The
  method also works for Weibull distribution as well as any distributions that
  can be transformed to location-scale families of distributions. Through simulation
  studies, we have demonstrated that the proposed LSE-GPQ method provide better
  coverage probabilities compared to traditional methods such as MLE-based
  asymptotic inference and parametric bootstrapping, especially in small sample scenarios.
  We have also applied the proposed method to real data examples, showing its practical
  utility in providing more accurate inference for the reliability function of the
  log--logistic distribution. 

  The construction of GPQ with LSE can be easily extended to the case of censored
  data, which is common in reliability inference and in survival analysis. Our simulation study showed that the proposed method still provides better coverage probabilities compared to traditional methods in small sample scenarios with light or moderate censoring.
  Additionally, in addition to the reliability function, the GPQs for the
  parameters of the log--logistic distribution can be easily mapped to the GPQ
  for any functions of the log--logistic parameters $\alpha$ and $\beta$, which allows
  us to construct confidence intervals for other quantities of interest.

  In the future, further research is warranted to investigate the performance of the proposed framework and explore its applications to other location–scale family distributions. Additionally, it would
  be interesting to investigate the ANOVA problems with log--logistic distribution
  assumptions, such as comparing reliability or mean survival time for different
  groups.

  \bibliographystyle{unsrt}
  \bibliography{references} 
\end{document}